\documentclass[10pt]{article}
\usepackage{latexsym}
\usepackage{caption}
\usepackage{graphics}
\usepackage{xcolor}
\usepackage{hyperref}
\usepackage{ulem}
\usepackage[affil-it]{authblk}
\usepackage{geometry}
\begin{document}
\title{Confinement versus interface bound states in spin-orbit coupled nanowires}


\author{Lorenzo Rossi, Fabrizio Dolcini, Fausto Rossi}

%
%
\affil{Dipartimento di scienza Applicata e Tecnologia, Politecnico di Torino, Torino, Italy}
%
%

\maketitle

\abstract{
Semiconductor nanowires with strong Rashba spin-orbit coupling are currently on the spotlight of several research fields  such as spintronics, topological materials and quantum computation. While most theoretical models  assume an infinitely long nanowire, in actual experimental  setups the nanowire has a finite length, is contacted to metallic electrodes and is partly covered by gates. By taking these effects into account through an inhomogeneous spin-orbit coupling profile, we show that in general two types of bound states arise in the nanowire, namely confinement bound states and interface bound states. The appearance of confinement bound states, related to  the finite length of the nanowire, is favoured by   a mismatch of the bulk band bottoms characterizing  the lead and the nanowire, and occurs  even in the absence of magnetic field. In contrast, an interface bound states may only appear if a magnetic field   applied perpendicularly to the spin-orbit field direction   overcomes a critical value, and is favoured by an alignment of the band bottoms of the two regions across the interface. We describe in details the emergence of these two types of bound states, pointing out their differences.   Furthermore, we show that  when a nanowire portion is covered by a gate  the application of a magnetic field can change the nature of  the electronic ground state from a confinement to an interface bound state, determining a redistribution of the electron charge. 
%
%
%

\section{Introduction}
\label{intro}
Bound states  play a relevant role in nanotechnological applications. For instance, it has been known for decades that a suitable engineering of  semiconductor heterostructures yields nanometer scale confined bound states  along the growth direction, forcing the   electron dynamics to effectively take  place in a two-dimensional plane, thereby creating a 2DEG. Also, since bound states are characterized by discrete energy separations that can even be greater than thermal energy, one can exploit them to realize optical devices, such as photodetectors or lasers. In the field of hetero-junctions, the existence of interface bound states at the   separation between two materials can reduce or even mask the desired features of the current-voltage characteristics, so that suitable techniques  such as lattice matching  have to be adopted to prevent their formation.

However, bound states are also crucial  in terms of fundamental Physics. In Condensed Matter Physics, for instance, it has been  realized that bound states 
can be the hallmark of topological transitions: When a material enters a topological phase, a bound state emerges at the interface with a  topologically trivial material\cite{kane-review,zhang-review,ando-review}. The Su-Schrieffer-Heeger model for trans-polyacetylene, for instance, describes a one-dimensional topological insulator, which exhibits localized bound states at the two ends of the chain when in the topological phase\cite{SSH_PRL,SSH_PRB,kane-lubensky_NaturePhys_2014,meier_NatureComm_2016}. Even more strikingly, at the boundaries of a topological superconductor, peculiar bound states have been predicted to emerge, which are equal to their anti-particles and are thereby called Majorana quasi-particles\cite{kitaev,alicea_review,fujimoto,aguado_review}.  Due to their exotic braiding properties and their robustness to decoherence effects, they are considered a promising platform for quantum computing\cite{freedman_RMP_2008,alicea_PRX_2016,dassarma-stanescu,akhmerov_2018,kouwenhoven-review}.\\

The huge  advances in the analysis of topological materials  has also renewed the interest of the scientific community in  the spin-orbit coupling. Such relativistic effect, which opened up in the 90s the way to spintronics\cite{datta-das_1990,nitta_APL_1999,awschalom_science_2001,dassarma_RMP_2004,awschalom_nature_2007,dePicciotto_2010}, is nowadays on the spotlight in the search for innovative topological materials\cite{bercioux-lucignano}. 
Indeed spin-orbit is for instance  the mechanism  underlying the appearance of topological helical edge states in quantum Spin Hall effect\cite{kane-mele2005a,kane-mele2005b,bernevig_science_2006,zhang_2008,molenkamp-zhang}. 
Also, when a semiconductor is  proximitized by an ordinary $s$-wave superconducting pairing,   the spin-orbit coupling  gives rise to an effective $p$-wave superconducting pairing, necessary for the appearance of Majorana quasi-particles\cite{fu-kane_PRL_2008,beenakker-akhmerov_2008,vonoppen_2010,dassarma_2010,yazdani_PRB,crepin-trauzettel-dolcini_PRB_2014},   as observed in ferromagnetic atomic chains deposited   on a superconductor\cite{yazdani-science},  and in proximizited InSb and  InAs nanowires\cite{kouwenhoven_2012,liu_2012,heiblum_2012,xu_2012,defranceschi_2014,marcus_2016,marcus_science_2016,kouwenhoven_2018}. 
 Furthermore, the remarkable progress  in  gating techniques allows   a broad tunability of the spin-orbit coupling~\cite{gao_2012,slomski_NJP_2013,wimmer_2015,nygaard_2016,sasaki_2017,tokatly_PRB_2017,loss_2018,tsai_2018}, making previously unexplored regimes accessible now.\\

While early studies have focussed on the  topological bound states of proximitized spin-orbit coupled nanowires, more recent works have pointed out  that in the presence of a superconducting coupling both topological and trivial bound states may exist\cite{lee-law_PRL_2012,dassarma_PRR_2020,dassarma_PRB_2017,loss_EPJB_2015,cayao_PRB_2015,aguado_2016,trauzettel_2018,frolov_2019,ricco_PRB_2019,tewari_2019,cayao_PRL_2019,aguado-loss_2019}. 
Also, quite recently it has been realized that peculiar bound states can appear even when no superconducting coupling is present, if magnetic domains induce an inhomogeneous magnetic field on the nanowire\cite{ronetti-loss_2019},  similarly to the magnetic confinement effects predicted in other materials\cite{egger_PRB_2015,levy-yeyati_PRB_2015,richter_PRL_2020}.\\

In this paper, we focus on a spin-orbit coupled nanowire in its normal phase, {\it i.e.} without superconducting coupling, characterized by an {\it inhomogeneous} Rashba spin-orbit coupling (RSOC). Such inhomogeneities appear quite naturally not only because of disorder, but also when a clean nanowire   is contacted to metallic electrodes (leads) and/or when a portion of the nanowire is covered by a gate that locally changes its Structural Inversion Asymmetry (SIA). By considering also the presence of a magnetic field applied perpendicularly to the spin-orbit field direction, we are able to identify  two essentially different types of bound states, namely the confinement bound states, and the interface bound states. After introducing in Sec.\ref{sec:2} the model and the method, in Secs.\ref{sec:3} and  \ref{sec:4} we discuss in details the  origin and the differences of these two types of bound states. Then, in Sec.\ref{sec:5} we consider the case where a nanowire portion  covered by  a gate  acquires a locally different RSOC value, and we show how an applied magnetic field can change the  electronic ground state from a confinement to an interface bound state.  Finally, in Sec.\ref{sec:6} we draw our conclusions.

\section{The model and the method}
\label{sec:2}
We  consider a nanowire  along the $x$ direction  deposited on a substrate. Because of the  SIA  arising at the interface with the substrate, in the nanowire a Rashba spin-orbit ``magnetic" field arises, lying on the substrate plane, perpendicularly to the nanowire axis. We denote by $z$ such direction and by $\alpha$ its Rashba spin-orbit coupling (RSOC) constant. Furthermore,   the presence of local gates deposited above some portions of the nanowire, or of leads contacted to the nanowire, locally alters the SIA. These situations can thus be modeled by an inhomogeneous RSOC profile $\alpha(x)$.
If we denote by $\hat{\Psi}(x)=( \hat{\Psi}_\uparrow(x) \,,\,\hat{\Psi}_\downarrow(x) )^T$ the electron spinor field, where $\uparrow,\downarrow$ identify the spin projections along the spin-orbit  field direction $z$,  the Hamiltonian of the  system reads
\begin{equation}\label{Ham}
\hat{\mathcal{H}}  = \int \hat{\Psi}^\dagger(x)\,\left(\frac{p_x^2}{2 m^*} \sigma_0 -\frac{\left\{ \alpha(x) , p_x\right\}}{2\hbar}  \sigma_z\, \, - h_x  \sigma_x\right)\,  \hat{\Psi}(x)\,dx
\end{equation}
where
$p_x=-i\hbar \partial_x$ is the momentum operator, $m^*$ is the electron effective mass, 
$\sigma_0$ the $2 \times 2$ identity matrix, and $\sigma_x,\sigma_y,\sigma_z$ are the Pauli matrices. Furthermore, 
  $h_x$ describes the Zeeman energy related to an external magnetic field applied along the nanowire axis. Note that, since $p_x$ does not commute with the inhomogeneous RSOC profile $\alpha(x)$, the anti-commutator is needed\cite{sanchez_2006,sanchez_2008,sherman_2011,sherman_2013,lucignano_2015,sherman_2017,dolcini-rossi_PRB_2018}. For the homogeneous case the solution is straightforwardly obtained, whereas to treat the inhomogeneous case we applied an exact numerical diagonalization approach, as  we shall briefly illustrate here below.\\

\subsection{The homogeneous case}
Let us start by briefly recalling the well know case of a nanowire  with a homogeneous RSOC profile $\alpha(x)\equiv \alpha$. In such case the momentum $p_x$ trivially commutes with the uniform  $\alpha(x)$, the Hamiltonian can be diagonalized   by Fourier transform, and the eigenstates are labelled by the wavevector $k$. If the  magnetic field is absent, the problem is particularly simple, as it is diagonal is spin space: The  RSOC lifts the degeneracy of  spin-$\uparrow$ and spin-$\downarrow$ states, whose parabolic spectra get centered at   $k=\pm k_{SO}$ and lowered by the spin-orbit energy $E_{SO}=\hbar^2 k_{SO}^2/2 m^*$, where $k_{SO}=m^* |\alpha|/\hbar^2$ is the spin-orbit wavevector. When a magnetic field $h_x$ is applied, it causes the opening of a gap $2\Delta_Z$  between the two bands  $E_\pm(k)= \hbar^2 k^2/2 m^*\,\pm \sqrt{(\alpha k)^2+ \Delta_Z^2}$ of the spectrum,  where $\Delta_Z=|h_x|$ shall be called  the magnetic gap energy. Two regimes can be distinguished, namely i) the Zeeman-dominated regime ($E_{SO}<\Delta_Z/2$) where both bands have a minimum at $k=0$, and ii) the Rashba-dominated regime ($E_{SO}>\Delta_Z/2$), where the lower band exhibits a local maximum at $k=0$ and two  minima $E_-^{min}=-E_{SO}(1+\Delta_Z^2/4 E_{SO}^2)$   at $k=\pm k^{min}$, where $k^{min}= k_{SO} \sqrt{1-\Delta_Z^2/4 E_{SO}^2}$. Furthermore, the spin of the eigenstates tilts with varying the wavevector~$k$.

\subsection{The inhomogeneous case}
\label{sec:2:2}
The inhomogeneous case cannot be treated analytically in general. Except for the case  of a piecewise profile, where the solution can be constructed by matching homogeneous solutions with appropriate boundary conditions\cite{sanchez_2006,sanchez_2008,lorenzo_2020}, a numerical approach is needed to obtain the spectrum and the eigenfunctions.   To this purpose, denoting by $\Omega$ the   length of the whole system  and imposing periodic boundary conditions over $\Omega$, we rewrite the Hamiltonian as  
\begin{eqnarray}\label{H-k-pre}
\hat{\mathcal{H}}  =\sum_{k_1,k_2}\sum_{s_1,s_2=\uparrow,\downarrow}  \hat{c}^\dagger_{k_1,s_1} H_{k_1,s_1;k_2 s_2} \,\hat{c}^{}_{k_2,s_2} \quad,
\end{eqnarray} 
where $\hat{c}^{}_{k,s}$ (with $k = 2\pi n/\Omega$ and $s=\uparrow,\downarrow$) are the discrete Fourier mode operators of the electron field operator 
$\hat{\Psi}(x)= \Omega^{-1/2} \sum_k   e^{i k x}  ( \hat{c}_{k\uparrow} , \hat{c}_{k\downarrow} )^T$, and 
\begin{eqnarray} \label{H-k-inhomo}
H_{k_1,s_1;k_2 s_2} = \left[ \left( \varepsilon^0_{k_1}\sigma_0- h_x \sigma_x \right) \delta_{k_1,k_2} \, - 
\alpha_{k_1-k_2}\frac{k_1+k_2}{2}\, \sigma_z\right]_{s_1,s_2}  \quad.
\end{eqnarray}
Here $\alpha_q$ is the (discretized) Fourier transform of the RSOC profile $\alpha(x)$.
An exact numerical  diagonalization of the Hamiltonian matrix Eq.(\ref{H-k-inhomo})  enables us to obtain the set $E_\xi$ of eigenvalues and the matrix $U$  of its eigenvectors. Then,   the original Fourier mode operators can be rewritten as $\hat{c}_a =\sum_\xi U_{a, \xi} \,\hat{d}_\xi$, where $a=(k,s)$ is a compact quantum number notation for the original basis and $\hat{d}_\xi$ are the 
diagonalizing operators, while the system Hamiltonian can be rewritten as
$ 
\hat{\mathcal{H}}=\sum_\xi E_\xi\, \hat{d}^\dagger_\xi \hat{d}^{}_\xi
$. By re-expressing  the electron field operator $\Psi_s(x)$ with spin component $s=\uparrow,\downarrow$  as 
$
\hat{\Psi}_s(x)=\Omega^{-1/2}\sum_{k,\xi} e^{i k x} U_{ks, \xi} \, \hat{d}_\xi$ and
by  exploiting $\langle \hat{d}^\dagger_\xi \hat{d}^{}_{\xi^\prime} \rangle_\circ=\delta_{\xi \xi^\prime} f^\circ(E_\xi)$, with $f^\circ(E)=\left\{1+\exp\left[(E-\mu)/k_B T\right]\right\}^{-1}$ denoting the Fermi distribution function, the equilibrium expectation value of the density operator 
\begin{equation}\label{charge-density-def}
\rho(x)=\left\langle \hat{\Psi}^\dagger (x)  \hat{\Psi}^{} (x) \right\rangle_\circ 
\end{equation}
 can be straightforwardly evaluated   as
  $\rho(x)=\sum_{\xi} \rho_\xi(x)$, where 
\begin{equation}
 \rho_\xi(x) = \frac{1}{L} \sum_{s=\uparrow,\downarrow} \sum_{k_1, k_2} e^{-i (k_1-k_2) x} \, U^*_{k_1 s,\xi} U^{}_{k_2 s,\xi}\, f^\circ(E_\xi)
\end{equation}
is the contribution arising from the $\xi$-th eigenstate.
In this way, the contribution of each eigenstate (in particular the bound state) can be singled out.

\section{Confinement bound states}
\label{sec:3}
In order to illustrate the emergence of confinement bound states, it is sufficient to consider the case without magnetic field ($h_x=0$). In this case, the Hamiltonian in Eq.(\ref{Ham}) is diagonal in spin space and, by performing the spin-dependent gauge transformation
\begin{equation}\label{gauge-transf}
\hat{\Psi}(x)=\displaystyle e^{i \frac{m^*}{\hbar^2}  \sigma_3\int^x_0 \alpha(x^\prime) dx^\prime} \, \hat{\Psi}^\prime(x)\quad,
\end{equation}
it can be rewritten  as
\begin{eqnarray} \label{Hameff}
\hat{\mathcal{H}}  =\int {\hat{\Psi}}^{\prime \dagger}(x)\, \left(\frac{p^2_x}{2 m^*}   + U_{SO}(x) \right) {\hat{\Psi}^\prime}(x) \,dx    \quad,
\end{eqnarray} 
where the effective potential 
\begin{equation}\label{ESO(x)}
U_{SO}(x)= -E_{SO}(x)=- \frac{{m^*}  \alpha^2(x)}{2\hbar^2}
\end{equation}
depending on the RSOC profile $\alpha(x)$ corresponds to (minus) the inhomogeneous Rashba spin-orbit energy. Notice that, due to the absence of magnetic field $h_x$,  the problem becomes purely scalar  when rewritten in terms of the new fields ${\hat{\Psi}^\prime}=(    \Psi^\prime_\uparrow \,,\,\Psi^\prime_\downarrow  )^T$. In terms of the original fields $\hat{\Psi}(x)=( \hat{\Psi}_\uparrow(x) \,,\,\hat{\Psi}_\downarrow(x) )^T$, the spin-$\uparrow$ and spin-$\downarrow$ components acquire opposite  space-dependent phase factors, as shown by Eq.(\ref{gauge-transf}). As an example, for a uniform RSOC $\alpha(x) \equiv \alpha$, one has 
\begin{equation}
\Psi_{\uparrow,\downarrow} (x) = e^{\pm i \, \mbox{\small sgn}(\alpha) \, k_{SO} x}\,\Psi^\prime(x)
\end{equation}
which corresponds, in momentum space, to   shifting horizontally  the parabolic spectrum by a spin-orbit wavevector $k_{SO}=m^*|\alpha|/\hbar^2$, in opposite directions for spin $s=\uparrow,\downarrow$.\\

\begin{figure}
\centering
\resizebox{0.7\textwidth}{!}{
\includegraphics{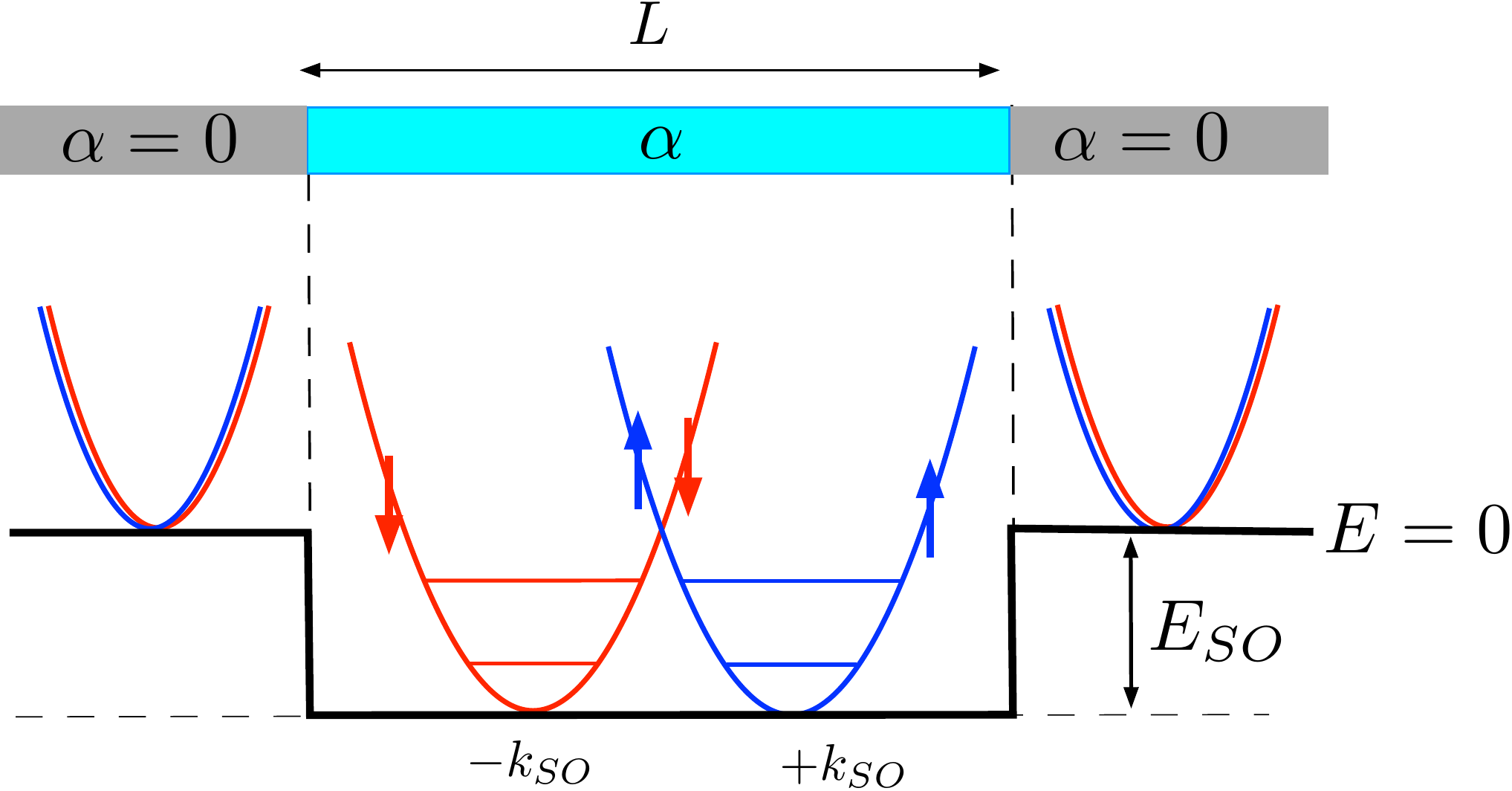}
}
\caption{Sketch of a nanowire coupled to two metallic leads in the absence of magnetic field, and of the related energy bands characterizing the bulks of the outer leads and of the nanowire. 
While  the leads are characterized by a vanishing RSOC and by a spin-degenerate parabolic spectrum, the RSOC $\alpha$ present in the  nanowire lifts the spin degeneracy  even without magnetic field. Furthermore, the energy bands are lowered by an amount corresponding to the spin-orbit energy $E_{SO}=m^* \alpha^2/2 \hbar^2$, giving rise to the potential well described by Eq.(\ref{ESO(x)}) and depicted by the thick black line. The finite length of the central nanowire yields the presence of confinement bound states, whose energy lie in the energy window between the band bottoms of the leads and the nanowire.} 
\label{fig:1}      
\end{figure}

\begin{figure}
\centering
\resizebox{0.9\textwidth}{!}{
\includegraphics{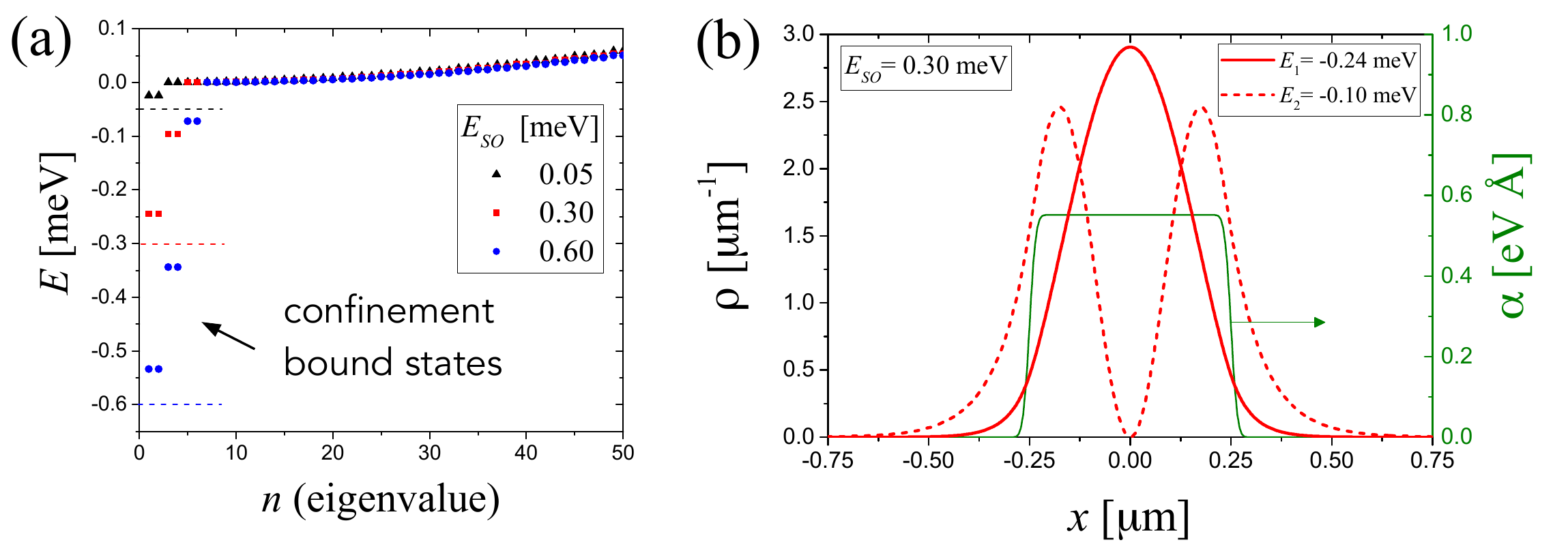}
}
\caption{Panel (a): The  energy spectrum of a {\rm InSb} nanowire+leads system, sketched in Fig.\ref{fig:1} and described by the inhomogeneous RSOC Eq.(\ref{alpha-double}), with a nanowire length   $L=500\,{\rm nm}$ and a smoothening length $\lambda_s=50\,{\rm nm}$. The effective mass is $m^*=0.015 m_e$. No magnetic field is applied ($\Delta_Z=0$). Different colors and symbols refer to three different values of the spin-orbit energy of the nanowire, $E_{SO}=0.05\,{\rm meV}$ (black triangles), $E_{SO}=0.30\,{\rm meV}$ (red squares), $E_{SO}=0.60\,{\rm meV}$ (blue circles).  Besides the continuum spectrum, discrete bound states appear, in spin-degenerate pairs, in the energy window between the bulk band bottom $E=0$ of the outer leads and the bulk band bottom $-E_{SO}$ of the nanowire (indicated by an horizontal dashed lines as a guide to the eye). Panel (b): for the case $E_{SO}=0.30\,{\rm meV}$, the spatial profiles of the density $\rho(x)$ of the ground bound state (solid red curve) and the first excited bound state (dashed red curve) are shown. The thin green curve displays the inhomogeneous spatial profile $\alpha(x)$ in Eq.(\ref{alpha-double}).} 
\label{fig:2}      
\end{figure}

For suitable inhomogeneous $\alpha(x)$ profiles, a possibility opens up that the effective potential Eq.(\ref{ESO(x)}) represents a quantum well hosting confinement bound states. This occurs, for instance, when a nanowire  characterized by a RSOC~$\alpha$ in its bulk  is   sandwiched between two metallic electrodes with vanishing RSOC, as sketched in Fig.\ref{fig:1}.   The simplest   model  describing this situation is  a square profile, $\alpha(x)=\theta( {L}/{2}-|x|)$, with $\theta$ denoting the Heaviside function. Then,  Eq.(\ref{ESO(x)}) represents a square quantum well with a width ${L}$ and a depth $-E_{SO}$ given by the bulk value of spin-orbit energy $E_{SO}=m^* \alpha^2/2 \hbar^2$ of the nanowire. 
As is well known, at least one bound state is always present, and the number of bound states increases with the magnitude of the RSOC in the nanowire. Furthermore, if the nanowire length ${L}$ is short enough, the energy separation between the   bound states becomes appreciable  (see Fig.\ref{fig:1}). 
\\A more realistic model to describe the nanowire+leads system  assumes a smoothened profile of the RSOC
\begin{eqnarray} \label{alpha-double}
\alpha(x)= \frac{\alpha}{2} 
  \left[{\rm Erf}\left(\frac{\sqrt{8}}{\lambda_s}(x+\frac{{L}}{2}) \right) -{\rm Erf}\left(\frac{\sqrt{8}}{\lambda_s}(x-\frac{{L}}{2})\right) \right] \quad, 
\end{eqnarray}
which goes from zero (leads) to $\alpha$ (nanowire bulk) within 2\% over a smoothening length $\lambda_s$. In Fig.\ref{fig:2} we analyze this case for a $500 {\rm nm}$ long {\rm InSb} nanowire (effective mass   $m^*=0.015 m_e$) contacted to two metallic electrodes, and for a smoothening length $\lambda_s=50\,{\rm nm}$, for three different values of RSOC corresponding to three different values of spin-orbit energy $E_{SO}$. Panel (a) displays the spectrum,  which exhibits both a continuum branch  for energies  above the band bottom $E=0$ of the outer leads, and some additional discrete bound states, always appearing in spin-degenerate pairs, whose number increases with the magnitude of the RSOC.   As expected, the bound states energies $E_{bs}$ are located in the energy window $-E_{SO} \le E_{bs} < 0$ between the bulk band bottom $-E_{SO}$ of the nanowire (indicated by dashed horizontal lines as a guide to the eye) and the bulk band bottom $E=0$ of the leads, as also sketched in Fig.\ref{fig:1}.  Figure \ref{fig:2}(b) shows, for the case $E_{SO}=0.30\,{\rm meV}$, the spatial profile of the density $\rho(x)$ of the ground bound state (solid red curve) and the first excited bound state (dashed red curve), as well as the inhomogeneous spatial profile $\alpha(x)$ (thin green curve).\\

We conclude this section by emphasizing once more that a prerequisite for the formation of a confinement bound state is that the RSOC profile varies non-monotonically. In  the case {\it e.g.} of one single interface separating two regions characterized by different RSOC values, where the profile $\alpha(x)$ varies {\it monotonically}  from the value $\alpha_L$ on the left of the interface to the value $\alpha_R$ on the right, the effective potential in Eq.(\ref{ESO(x)}) never creates a quantum well. Indeed, if $\alpha_L$ and $\alpha_R$ have the same sign, $U_{SO}$ also changes monotonically, whereas if $\alpha_L$ and $\alpha_R$ have opposite signs, so that the profile $\alpha(x)$ crosses zero, $U_{SO}$ describes a {\it barrier} at the interface. In neither case a  monotonic $\alpha(x)$ profile can give rise to bound states. This means that  {\it no interface bound state} is present, without magnetic field.  As we shall see in next Section, the situation is different when a magnetic field is applied.

\section{Interface bound states}
\label{sec:4}
When a magnetic field $h_x$ is applied along the nanowire axis, {\it i.e.}, perpendicularly to the spin-orbit field, another type of bound states can emerge when the RSOC profile $\alpha(x)$ is  inhomogeneous. Before discussing the formation of such bound state, we wish to point out that the inhomogeneous RSOC problem in the presence of an applied magnetic field is intrinsically more difficult than the field-free case. To illustrate that, we  apply again   the gauge transformation (\ref{gauge-transf}), and rewrite the Hamiltonian (\ref{Ham})   as
\begin{eqnarray} \label{Hameff_2}
\hat{\mathcal{H}}  =\int {\hat{\Psi}}^{\prime \dagger}(x)\, \left(\frac{p^2_x}{2 m^*}   + U_{SO}(x) -h_x \left( \sigma_x \cos\theta_{SO}(x)+\sigma_y \sin\theta_{SO}(x) \right)\right) {\hat{\Psi}^\prime}(x) \,dx    \quad,
\end{eqnarray} 
where $\theta_{SO}(x)=2m^*\int_0^x \alpha(x^\prime) dx^\prime /\hbar^2$ is called the spin-orbit angle.  In terms of the new fields ${\hat{\Psi}^\prime}$ the RSOC has been re-absorbed into  the previously discussed potential Eq.(\ref{ESO(x)}),  whereas  the original  uniform magnetic field has  transformed  into 
an effective {\it inhomogeneous} magnetic field, whose effects are more subtle. Still, the problem can be attacked, without even performing the  gauge transformation, by the method described in Sec.\ref{sec:2:2}. The results, which we shall now illustrate here below,  show  the emergence of interface bound states.\\

Differently from   confinement bound states,  the interface bound states can emerge even when the RSOC profile varies monotonically across one single interface  from a value $\alpha_L$ (on the left)   to $\alpha_R$ (on the right),   over a lengthscale~$\lambda_s$, 
\begin{equation}\label{alpha(x)}
\alpha(x)=\frac{\alpha_R+\alpha_L}{2}+\frac{\alpha_R-\alpha_L}{2}\, \mathrm{Erf}\left(\frac{\sqrt{8} \,x }{ \lambda_{s}}\right) \quad,
\end{equation}
where we have located the interface at $x=0$ without loss of generality.   
It turns out that the formation of interface bound state is particularly favorable when the sign of the RSOC changes across the interface, as can be achieved by appropriate gating techniques\cite{slomski_NJP_2013,kaindl_2005,wang-fu_2016,nitta-frustaglia}.
To illustrate such effect,  we shall thus focus on the case where the RSOC changes  from $\alpha_L=\alpha>0$ to $\alpha_R=-\alpha<0$.   
Notice that, since the spin-orbit energy $E_{SO}=m^* \alpha^2/2 \hbar^2$ depends on the square of the RSOC, the bulk band bottoms   take  the {\it same} values on both sides of the interface, as sketched in Fig.\ref{fig:3}.  In  Fig.\ref{fig:4}(a) the energy spectrum  is explicitly shown for such interface with  smoothening length $\lambda_s=50\,{\rm nm}$ in  a {\rm InSb} nanowire,  for the case of spin-orbit energy   $E_{SO}=0.50\,{\rm meV}$, and for three different values of the magnetic gap energy $\Delta_Z=0$ (black triangles), $\Delta_Z=0.5 \,{\rm meV}$ (red squares) and $\Delta_Z=1.0 \,{\rm meV}$ (blue circles). While the spectrum is purely continuous for vanishing magnetic field, when $\Delta_Z>0$ one single bound state appears.  For the latter two positive values of applied magnetic field, the density profile of the bound states is plotted in Fig.\ref{fig:4}(b), showing that the bound state is located at the interface.  

\begin{figure}
\centering
\resizebox{0.7\textwidth}{!}{
\includegraphics{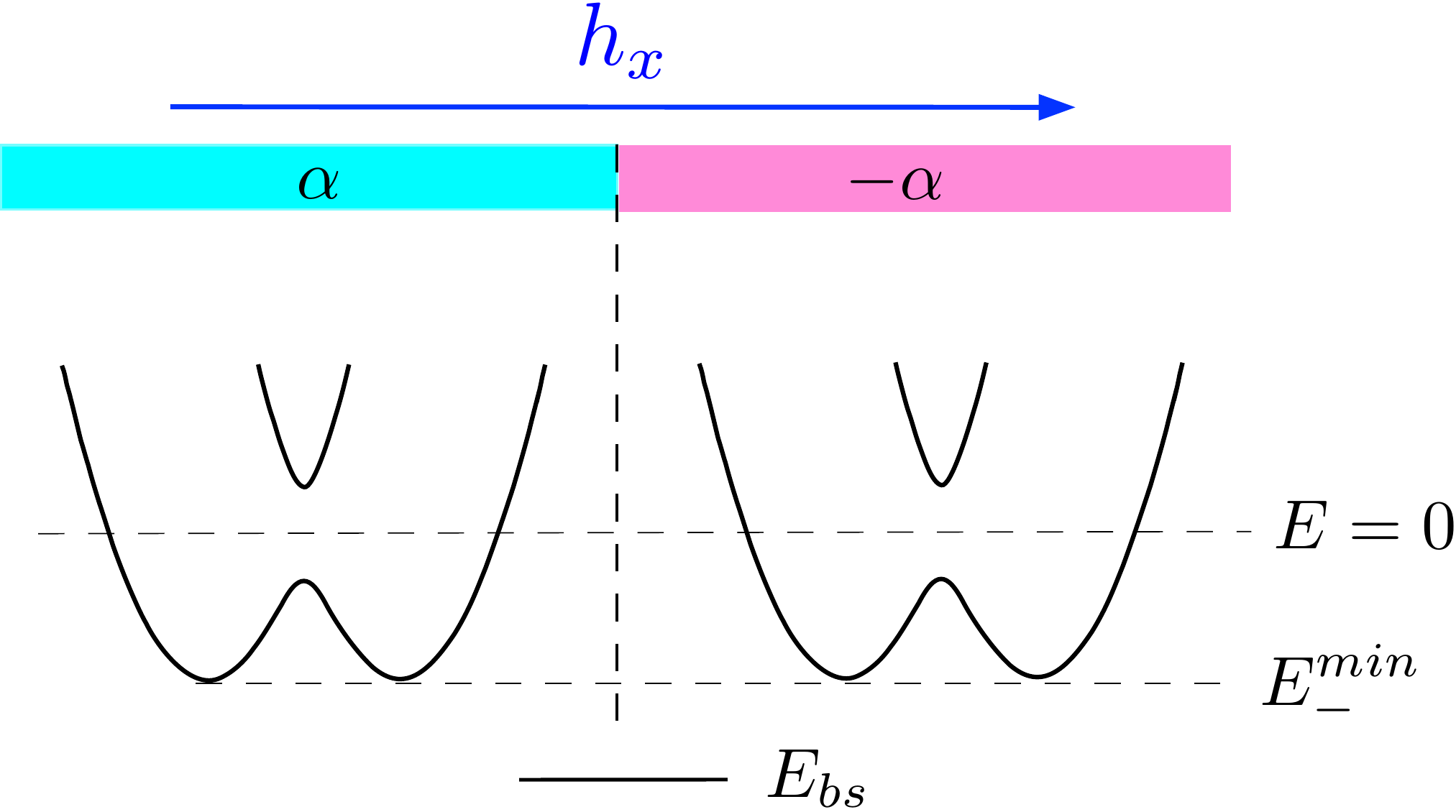}
}
\caption{Sketch of an interface between two portions of a nanowire characterized by two different values of RSOC, in the presence of an applied magnetic field along the nanowire axis. In particular, when the RSOC takes equal and opposite values across the interface, the spin-orbit energy of the two sides, which depends on the square of $\alpha$, is the same, so that the two bulk bands exhibit the same spectrum and their band bottoms are aligned. However, an interface bound state appears, energetically located {\it below} the continuum spectrum.} 
\label{fig:3}      
\end{figure}

\begin{figure}
\centering
\resizebox{0.9\textwidth}{!}{
\includegraphics{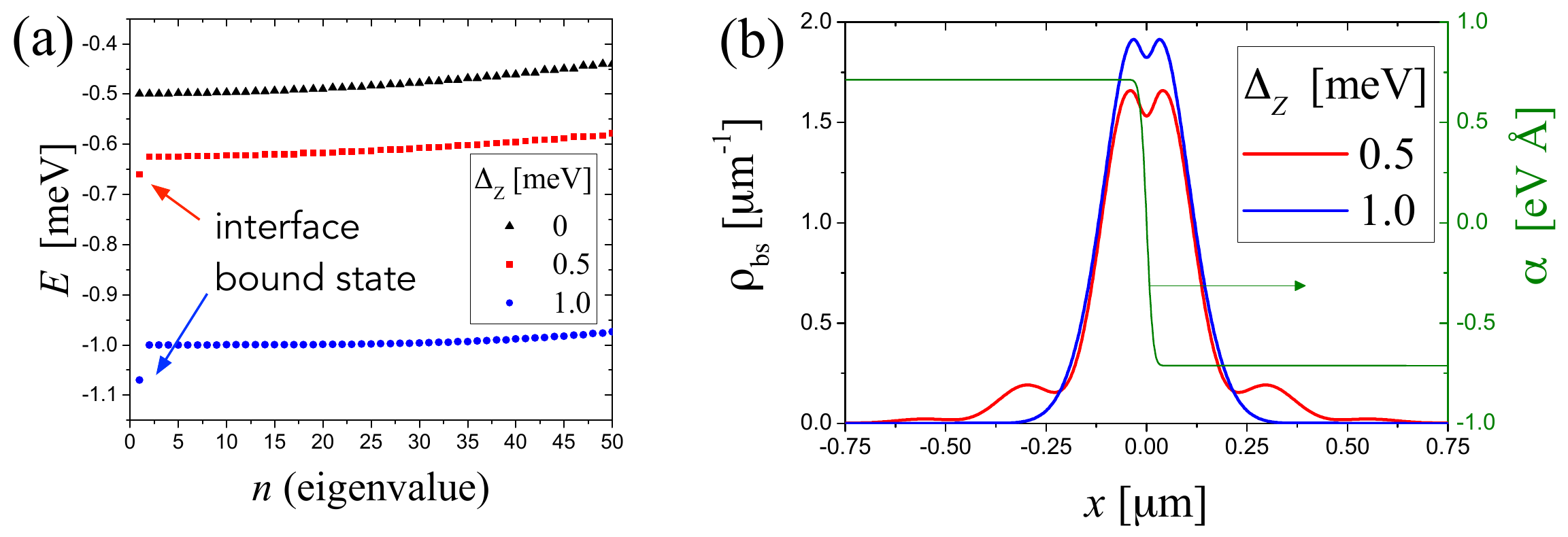}
}
\caption{Panel (a): Energy spectrum of inhomogeneous RSOC profile (\ref{alpha(x)}) describing the interface sketched in Fig.\ref{fig:3}, i.e. $\alpha_L=\alpha$ and $\alpha_R=-\alpha$, with smoothening length $\lambda_s=50\,{\rm nm}$, in a {\rm InSb} nanowire ($m^*=0.015 m_e$). The spin-orbit energy characterizing both sides is $E_{SO}=0.5 \,{\rm meV}$ and the three different curves refer to three different values of the  magnetic gap energy $\Delta_Z=|h_x|$. While for vanishing magnetic field the spectrum has a purely continuum branch, for non-vanishing magnetic field a bound state appears below the continuum branch. Panel (b): the density profile of the bound state is plotted for the two non-vanishing magnetic field values, showing that the bound state is located at the interface. The thin green curves describes the RSOC profile Eq.(\ref{alpha(x)}).}  
\label{fig:4}      
\end{figure}

Two further differences from the confinement bound states are noteworthy. First the energy of the interface state  lies {\it below} the band bottoms of the two regions, corresponding to the bottom of the continuum branch of the spectrum, as is clear from Fig.\ref{fig:3}. Second, its appearance is mostly favoured by an alignment --rather than a mismatch-- of the bulk band bottoms  across the interface, as in the   case illustrated in Fig.\ref{fig:3} and analyzed in Fig.\ref{fig:4}. 
We mention  that, when the two spin-orbit energies across the interface are different, the alignment can be restored if  the applied magnetic is sufficiently strong. Then, the interface bound state appears for magnetic gap energy above a threshold value $\Delta_Z^\star$,  as has been recently shown\cite{lorenzo_2020}.

Before concluding this section, it is  worth recalling that other types of bound states may appear at the interface between two different materials. As is well known, interface states may be caused by the band bending  near the interface, which is particularly relevant in metal-semiconductor junctions, where the  difference between the Fermi energies is large. Those interface states thus have an essentially electrostatic origin. In striking contrast, the interface bound state  described here only exists if a magnetic field is applied and only if RSOC is present, as pointed out at the end of Sec.\ref{sec:3}. Its emergence is thus an essentially magnetic and spinorial effect. Indeed it can also exist if the band spectra across the interface are perfectly equal  (see Figs.\ref{fig:3} and \ref{fig:4}), i.e. where the customary interface bound states are not expected to exist. Of course, in a more general situation also the electrostatic interface bound states can in principle be present.

\section{Nanowire covered by a gate and exposed to a magnetic field}
\label{sec:5}
Let us now consider the case where a gate partly covers   the nanowire, thereby locally changing the SIA and   the RSOC  of the nanowire region underneath,     similarly to what occurs in constrictions in quantum spin Hall systems\cite{sassetti-citro_2012,sternativo_2014,molenkamp-buhmann_2020}.
This situation,  sketched in Fig.\ref{fig:5}(a), can be described by a RSOC profile
\begin{eqnarray}\label{alpha-double-2}
\alpha(x) = \alpha_{out}+\frac{\alpha_{in}-\alpha_{out}}{2}   \left[{\rm Erf}\left(\frac{\sqrt{8}}{\lambda_s}(x+\frac{L}{2}) \right) -{\rm Erf}\left(\frac{\sqrt{8}}{\lambda_s}(x-\frac{L}{2})\right) \right] \quad, 
\end{eqnarray}
where  $L$ is the length of  the central region, the origin $x=0$ is set  in its middle point, and $\alpha_{in}$ and $\alpha_{out}$ denote the bulk RSOC values   of the central region and outer regions, respectively.  
For definiteness, we shall focus on the situation $|\alpha_{in}|>|\alpha_{out}|$, which generalizes the case $\alpha_{out}=0$ of the metallic leads discussed in Sec.\ref{sec:3}.  Furthermore, since the band bottom of the central region is already modified indirectly through the RSOC, we shall neglect here the change induced directly by the gate voltage, which only involves the charge and has no effect on the spin degree of freedom.

We shall analyze  the spectrum of such inhomogeneous system and, in particular, we shall discuss how it is modified when a magnetic field is applied. 
Based on the material previously discussed in Sec.\ref{sec:3} and  Sec.\ref{sec:4}, let us first point out the scenario one can expect in this situation. On the one hand, when no magnetic field is present, the band bottom $-E_{SO,in}$ of the central region is lower than the outer band bottom $-E_{SO,out}$, and confinement bound states  exist, while no interface bound state may be present. On the other hand, when a magnetic field is applied,  the confinement bound states  get modified by the magnetic field, while additional interface bound states  appear at the two interfaces. The latter are energetically located below the lower bulk band bottom and are  thus  more favorable than confinement bound states. In fact, when the magnetic field is sufficiently strong, the band bottoms of the central and outer regions get aligned and the confinement bound states  disappear completely, leaving only the interface bound states.\\

\begin{figure}
\centering
\resizebox{\textwidth}{!}{
\includegraphics{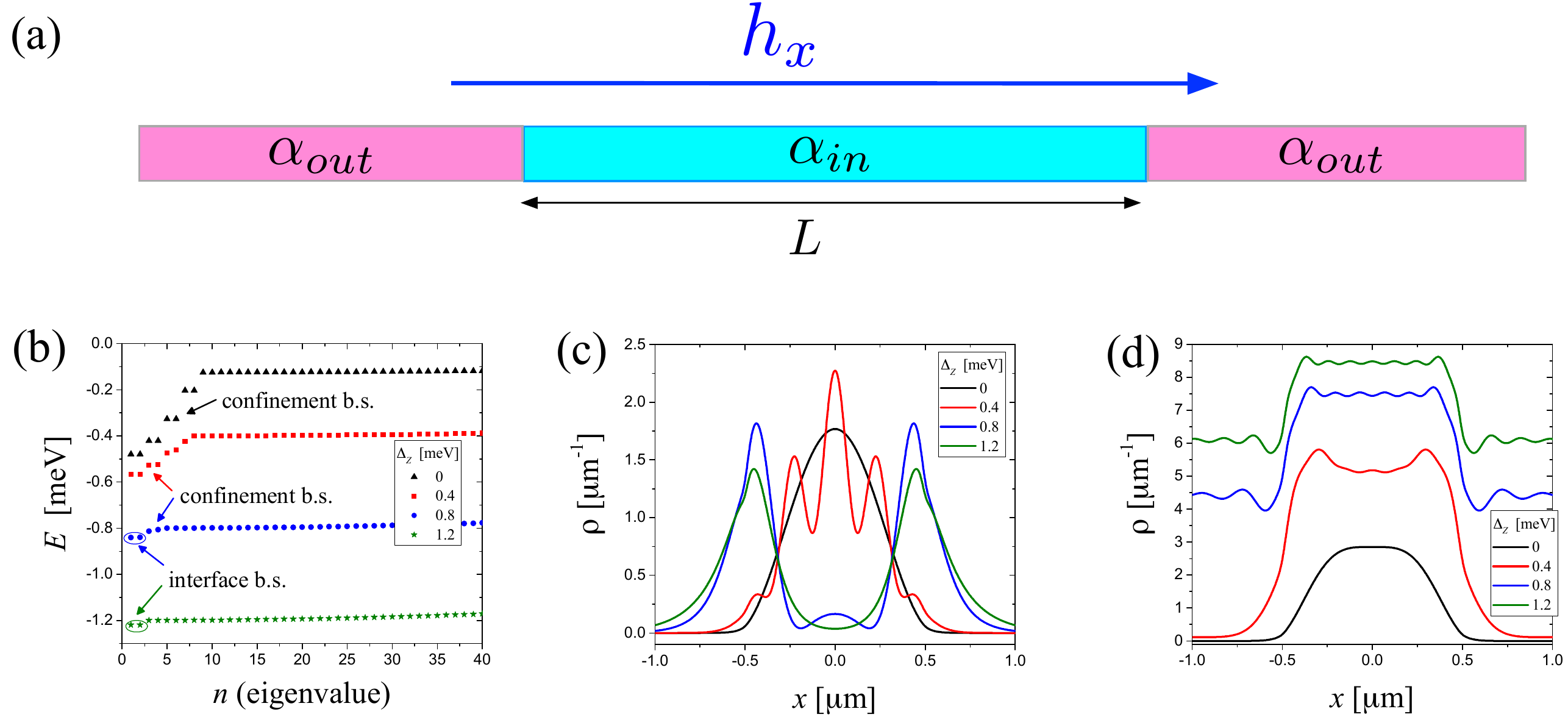}
}
\caption{Panel (a): A  $1 \mu{\rm m}$-long portion of a {\rm InSb} nanowire (effective mass  $m^*=0.015 m_e$) takes a RSOC value $\alpha_{in}$  different  from the value $\alpha_{out}$ characterizing the rest of the nanowire,  {\it e.g.} due to the presence of a metallic gate covering it.  The RSOC $\alpha_{in}$ corresponds to a bulk spin-orbit energy $E_{SO,in}=0.5 \,{\rm meV}$ for the central region, while $\alpha_{out}=-\alpha_{in}/2$, and $E_{SO,out}=0.125 \,{\rm meV}$. The smoothening length of the RSOC profile (\ref{alpha-double-2}) is $\lambda_s=50\,{\rm nm}$. The inhomogeneous nanowire is exposed to an external magnetic field along the nanowire axis. Panel (b): The spectrum of the inhomogeneous nanowire    is plotted for four different values of the applied magnetic field:  for vanishing or weak magnetic field (black triangles and red squares) only  confinement  bound states are present. For $\Delta_Z>\Delta_Z^\star \simeq 0.5 \,{\rm meV}$,  two additional interface bound states appear  below the confinement bound states  (blue circles), while for  $\Delta_Z>2 E_{SO}=1\,{\rm meV}$ (green stars) the confinement bound states have disappeared and only the interface bound states survive. Panel (c): The density profile $\rho_{lowest}$ of the lowest energy state, plotted for the same four values of applied magnetic field, shows a change in the nature of the electronic ground state   from a confinement to   interface bound states, determining a displacement of the electron charge from the center to the interfaces with the leads located at $x=\pm 0.5\,{\rm \mu m}$.   Panel (d): the profile of the total density $\rho$, involving all occupied states up to a chemical potential $\mu=-0.45\,{\rm meV}$ is shown for the four different values of applied magnetic field.} 
\label{fig:5}      
\end{figure}
We   illustrate   these effects in a  {\rm InSb} nanowire, where the central region has a  bulk spin-orbit energy     $E_{SO,in}=0.5 \, {\rm meV}$, while the outer regions is characterized by $\alpha_{out}=-\alpha_{in}/2$, yielding $E_{SO,out}=0.125 \, {\rm meV}$. The system in Fig.\ref{fig:5}(a) is modeled  by an inhomogeneous RSOC profile Eq.(\ref{alpha-double-2}), where the length of the central region is ${L}=1\,\mu{\rm m}$ and the smoothening length across each interface is $\lambda_s=50\,{\rm nm}$.    In  Fig.{\ref{fig:5}(b) the spectrum of the inhomogeneous nanowire is plotted for four different values of the applied magnetic gap energy $\Delta_Z=(0,0.4,0.8,1.2)\, {\rm meV}$. As one can see, for vanishing magnetic field $\Delta_Z=0$ (black triangles), four doubly degenerate confinement bound states are present, within the energy window between the band bottoms $-E_{SO,out}=-0.125\,{\rm meV}$ and $-E_{SO,in}=-0.50\,{\rm meV}$ of the outer and central regions, respectively. When the magnetic field is increased (red squares), the energy window determined by such band bottom mismatch   reduces, and so does the number of confinement bound states. Furthermore, if the magnetic gap energy overcomes a threshold value $\Delta_Z > \Delta_Z^\star  \simeq 0.5\,{\rm meV}$ (blue circles), two additional interface bound states appear. They are linear combinations of the two bound states appearing at the two interfaces  and are almost degenerate, with a tiny energy splitting caused by a non vanishing overlap due to the finite length $L$ of the central region. Note that in this situation confinement and interface bound states coexist, although the interface bound state are always energetically more favorable, as they lie below the band bottoms. However, for even stronger magnetic fields, $\Delta_Z \ge 2 E_{SO} =1\,{\rm meV}$, the confinement bound states disappear and only the interface bound states survive  (green stars). 

In Fig.\ref{fig:5}(c) we have plotted   the density profile $\rho_{lowest}$ of the  lowest energy state, for each of the four $\Delta_Z$ values. One can thus clearly see that, while for vanishing magnetic field (black curve) the energetically  most favorable state  is mainly located at the center of the nanowire, by increasing the magnetic field the interface bound state becomes more favorable (blue and green curves). By operating with the magnetic field one can thus displace the  charge of the electronic ground state from the center of the gated nanowire region towards the interfaces  located at $x=\pm 0.5\,\mu{\rm m}$,  yielding a stronger coupling with the outer regions, which play the role of leads. Finally, in Fig.\ref{fig:5}(d) we have plotted the full electron density, due to all states filled up to a chemical potential value $\mu= -0.45 \,{\rm meV}$, again for the four values of applied magnetic gap energy. While for  $\Delta_Z=0$ the charge is purely localized in the center of the nanowire, the application of a magnetic field leads the charge to be delocalized also in the outer `leads'. Notably, even for the green curve at $\Delta_Z=1.2\,{\rm meV}$, where both nanowire regions are in the Zeeman dominated regime ($\Delta_Z>2 E_{SO,in}> 2  E_{SO,out}$) and  their bulks have the same band bottom, the stronger spin-orbit coupling in the central region causes the density therein to exhibit   a plateau higher  than the density in the outer regions.

\section{Conclusions}
\label{sec:6}
In conclusion, in this paper we have investigated the presence of bound states in  spin-orbit coupled nanowires characterized by inhomogeneous RSOC profile. This can account for various effects, namely the finite length of the nanowire, the contacts to metallic leads or the situation  where the RSOC is locally modified by the presence of a gate  covering part of the nanowire. We have shown that two types  of bound states exist, namely the confinement bound states and the interface bound states, with quite different origin and features, which we can now summarize. 

The confinement bound states, described in  Sec.\ref{sec:3}, exist when a non-monotonic RSOC profile $\alpha(x)$ creates an effective confinement potential Eq.(\ref{ESO(x)}). The typical situation where this occurs is when a nanowire with finite length $L$ is contacted through two interfaces to two electrodes where the RSOC vanishes, where Eq.(\ref{ESO(x)}) represents a quantum well, with a depth given by the spin-orbit energy $E_{SO}$ of the nanowire and a width corresponding to the nanowire length~$L$. Their energies lie in the energy window between the bulk band bottom  of the   nanowire and bulk band bottom of the leads. The emergence of these states  is thus related to the {\it mismatch} of the two band bottoms. These states exist also when no magnetic field is applied, and the application of a magnetic tends to hinder their existence, since for a sufficiently strong magnetic field both the nanowire and the leads enter the Zeeman-dominated regime where the band bottom equals $-|h_x|$. 

In contrast, the  interface bound state  described in Sec.\ref{sec:4} is present only when the magnetic field $h_x$ (perpendicular to the spin-orbit field) is applied. It  may exist also for a  monotonic RSOC profile $\alpha(x)$, like in the presence of one single interface. This can be either a nanowire/lead interface or an interface between two  different portions of the nanowire, one being e.g. covered by a gate altering the  RSOC underneath, possibly with changing its sign. 
Differently from the confinement bound states, the existence of the interface bound states is favored by the {\it alignment} of the band bottoms of the two sides of the interface. For instance, when the RSOC takes equal and opposite  values across the interface, the two band bottom energies, which depend only on the square of the RSOC, align and these bound states exist for any weakly applied magnetic field. In general, for any two different bulk values of the RSOC $\alpha_L$ and $\alpha_R$ across the interface, like for a lead-interface, the interface bound state appear for a sufficiently strong magnetic field. In striking difference from the confinement bound states, the energy of the interface bound states lies {\it below} the bulk band bottoms. 

Finally, in Sec.\ref{sec:5}, we have considered the case where a nanowire portion acquires a locally stronger RSOC, {\it e.g.} due to a gate covering it. We have shown that, while for vanishing magnetic field the lowest energy state is a confinement bound state characterized by an electron density peaked at the center of the gated region, when a magnetic field is   increasingly applied the confinement bound states eventually disappear and the ground state consists of interface bound states. The ground state charge can thus be magnetically displaced towards the interfaces, leading to a stronger coupling to the outer regions, which play the role of leads. \\

The parameter values and the conditions  described above are at experimental reach in realistic setups with InSb and InAs nanowires\cite{kouwenhoven_2012,liu_2012,heiblum_2012,xu_2012,defranceschi_2014,marcus_2016,marcus_science_2016,kouwenhoven_2018}. The described bound states could possibly be exploited for photo-excitation in spin-orbit nanowires, similarly to what has been done with helical edge states of quantum Spin Hall effects\cite{kindermann2009,cayssol2012,artemenko2013,dolcetto-sassetti2014,artemenko2015,dolcini_PRB_2016,ganichev_PRB_2017,cayssol_PRB_2019}  
or for the analysis of out of equilibrium effects caused by a quench, as recently proposed\cite{cavaliere_PRB_2019}.

%

%
%

\end{document}